\documentclass[aps,prl,twocolumn,groupedaddress,showpacs]{revtex4}
\usepackage{graphicx}

\begin{document}

% Use the \preprint command to place your local institutional report
% number in the upper righthand corner of the title page in preprint mode.
% Multiple \preprint commands are allowed.
% Use the 'preprintnumbers' class option to override journal defaults
% to display numbers if necessary
%\preprint{}

%Title of paper
\title{Unexpected Effect of Internal Degrees of Freedom on Transverse Phonons in Supercooled Liquids.}

% repeat the \author .. \affiliation  etc. as needed
% \email, \thanks, \homepage, \altaffiliation all apply to the current
% author. Explanatory text should go in the []'s, actual e-mail
% address or url should go in the {}'s for \email and \homepage.
% Please use the appropriate macro foreach each type of information

% \affiliation command applies to all authors since the last
% \affiliation command. The \affiliation command should follow the
% other information
% \affiliation can be followed by \email, \homepage, \thanks as well.

\author{A. Patkowski}
\email[Author to whom corespondence should be addressed. Electronic mail: ]{patkowsk@amu.edu.pl}
%\homepage[]{Your web page}
%\thanks{}
%\altaffiliation{}

\author{J. Gapinski}
\affiliation{Institute of Physics, A. Mickiewicz University, Umultowska 85, 61-614, Poznan, Poland}

\author{G. Meier}
\author{H. Kriegs}
\affiliation{Institut für Festkörperforschung, Forschungszentrum Jülich, Postfach 1913, 52425 Jülich, Germany}

\author{A. Le Grand}
\author{C. Dreyfus}
\affiliation{P.M.C., U.M.R.7602, Case Postale 86, UPMC, 4 Place Jussieu, 75005 Paris, France}

%Collaboration name if desired (requires use of superscriptaddress
%option in \documentclass). \noaffiliation is required (may also be
%used with the \author command).
%\collaboration can be followed by \email, \homepage, \thanks as well.
%\collaboration{}
%\noaffiliation

\date{\today}

\begin{abstract}
We show experimentally that in a supercooled liquid composed of molecules with internal degrees of freedom the internal modes contribute to the frequency dependent shear viscosity and damping of transverse phonons, which results in an additional broadening of the transverse Brillouin lines. Earlier, only the effect of internal modes on the frequency dependent bulk viscosity and damping of longitudinal phonons was observed and explained theoretically in the limit of weak coupling of internal degrees of freedom to translational motion. A new theory is needed to describe this new effect.  We also demonstrate, that the contributions of structural relaxation and internal processes to the width of the Brillouin lines can be separated by measurements under high pressure.
\end{abstract}

% insert suggested PACS numbers in braces on next line
% insert suggested keywords - APS authors don't need to do this
%\keywords{}

%\maketitle must follow title, authors, abstract, \pacs, and \keywords
\maketitle

% body of paper here - Use proper section commands
% References should be done using the \cite, \ref, and \label commands
%\section{}
% Put \label in argument of \section for cross-referencing

Sound waves propagating in supercooled liquids are damped due to different relaxation processes in the medium. This damping results in broadening of the Brillouin lines observed in the light scattering Brillouin spectra. Damping of longitudinal phonons in glass forming liquids is usually related to structural relaxation processes and density fluctuations. The effect of internal molecular relaxations on phonon damping is much less studied and understood. 

Zwanzig \cite{Zwanzig} has shown that in the case of a liquid which consists of molecules with internal degrees of freedom, in the limit of very weak coupling of internal and centre of mass motions, the longitudinal phonons are broadened but the transverse phonons (and the shear viscosity) are unaffected by the internal relaxations. This result is sometimes interpreted by saying that the transverse phonons (and the shear viscosity) are never affected by internal relaxations. This feature, i.e. that a relaxation process contributes to damping of longitudinal, but not transverse phonons is sometimes used \cite{Monaco2} to assign this process to intramolecular dynamics. As we will show in this letter, this is correct in the case of weak coupling limit, but not in general. 

In order to understand better the theoretical results of Zwanzig \cite{Zwanzig} and Mountain \cite{Mountain} let us briefly recall what is meant by the limit of weak coupling of internal and translational dynamics, in the sense used by Zwanzig. The total Hamiltonian of the system consists of parts related to the centre of mass motions $H_C$, internal motions $H_I$ and interactions between these two sets of variables $\lambda U'$: $H = H_C + H_I + \lambda U'$, where $\lambda$ is the measure of these interactions. $H_C$ and $H_I$ consist, obviously, of the corresponding kinetic and potential energies of the centre of mass and internal motions, respectively. In the limit of weak coupling $\lambda$ is very small. In this limit the interaction energy $\lambda U'$ can be neglected and thus, the averages of functions as well as time derivatives can be separated in two independent (centre of mass and internal) parts \cite{Zwanzig}. Only in this limit Zwanzig has shown that the frequency dependent bulk viscosity and the width of longitudinal Brillouin lines depend on internal relaxations, while these relaxations do not contribute to the frequency dependent shear viscosity and therefore to the width of transverse Brillouin lines. 

To the best of our knowledge, the effect of clearly assigned internal relaxation on the damping of phonons in supercooled liquids has not been studied (systematically) as yet. The best documented case so far is ortho-terphenyl (OTP) where the relaxation processes as well as longitudinal and transverse phonons were studied in a broad temperature range. A fast process in OTP of a relaxation time of a few ps was found \cite{Steffen1, AP, Monaco1} and it was assigned to internal relaxation \cite{Monaco1, Monaco2}, mainly on the basis that it was contributing to the damping of  longitudinal but not transverse phonons, as predicted by Zwanzig \cite{Zwanzig}. 

A fast vibrational relaxation process of a similar relaxation time of a few picoseconds was also found in molecular dynamics simulations \cite{Mossa} and it was assigned to the phenyl - phenyl stretching. As an internal relaxation process, it can  contribute to the damping of the longitudinal (but not transverse) phonons in the weak coupling limit, as discussed by Zwanzig. Thus, temperature dependent experimental studies on OTP have shown, that in this liquid there exists fast internal dynamics that results in additional damping (broadening) of longitudinal but not transverse phonons \cite{Steffen1, AP, Monaco1, Monaco2}, in agreement with the theoretical model developed by Zwanzig. 

The question that should be asked at this point is, whether this limit of weak coupling can be generalized for all glass forming liquids, i.e. whether the internal relaxations are always weakly coupled to translational degrees of freedom, never contributing to the frequency dependent shear viscosity and to additional broadening (damping) of transverse Brillouin lines (phonons). As we show in this letter, such generalization is incorrect since there are cases when the internal relaxations also contribute to the frequency dependent shear viscosity and result in a substantial broadening of the transverse Brillouin lines.

In order to check the effect of internal modes on the width of longitudinal and transverse Brillouin lines (phonon damping) we have studied two molecules: 1,1'-bis(p-methoxyphenyl)cyclohexane (BMPC) and 1,1'-bis(4-methoxy-5-methylphenyl)cyclohexane (BMMPC). Their chemical structure is very similar (Fig.1) the only difference being the presence of an additional methyl group in the ortho position in both phenyl rings in BMMPC.

\begin{figure}
\includegraphics[angle=0,width=0.45\textwidth]{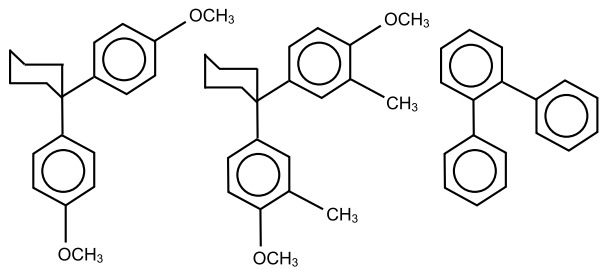}
\caption{\label{Fig.1}The chemical structures of BMPC (left), BMMPC (middle) and OTP (right) with indicated rotations of the phenyl ring.}
\end{figure}

This small chemical difference results in a dramatic difference in the dynamic processes observed in both molecules \cite{Gerd}: In BMPC the structural relaxation and two additional internal modes, related to the phenyl flip and rotation of OCH$_3$ group exist. The assignment of molecular origin of the internal modes has been obtained from deuterium NMR (phenyl flip) and Dielectric Spectroscopy (OCH$_3$ group rotation) studies. In BMMPC the presence of the additional methyl groups causes an almost complete freezing of both internal modes \cite{Gerd}. Thus, the pair of molecules: BMPC and BMMPC is very well suited for experimental studies of the effect of internal modes on the damping (width) of the longitudinal and transverse phonons (Brillouin lines). 

BMMPC and BMPC was prepared at the Max Planck Institute for Polymer Research, Mainz, Germany, according to the procedures described elsewhere \cite{Gerharz}. The glass transition temperature amounted to 265 and 245 K, respectively \cite{Gerharz, Gerd, Kahle}. The samples were filtered into the light scattering cells (Millipore filters, pore size 0.22 nm) at temperatures above their melting point. 

We have measured the Brillouin spectra of both BMPC and BMMPC in a broad temperature range. Additionally, pressure dependent studies were performed in order to separate the effects of pressure sensitive structural relaxation from that of pressure insensitive internal dynamics on the damping (width) of longitudinal and transverse phonons (Brillouin lines).

Brillouin spectra were measured using a 6-pass Tandem Fabry-Perot interferometer (Sandercock) and either Ar-ion laser (Spectra Physics, $\lambda = 514.5$ nm) or Nd:YAG laser (Coherent, $\lambda = 532$ nm) as a source of the incident light. The polarized (VV) spectra were measured in the back-scattering geometry, while for the depolarized (VH) spectra $90^o$ scattering angle geometry was used. For pressure dependent studies a high pressure cell operating in the pressure range of 1 - 2000 bar was used; the high pressure light scattering setup is described in detail elsewhere \cite{HPcell}. All presented values of the position $\omega_B$ and the half width at half height $\Gamma_B$ of the Brillouin lines have been recalculated for $\lambda = 514.5$ nm and a scattering angle of $90^{o}$. 

A typical set of polarized (VV) Brillouin spectra measured in the back scattering geometry for BMMPC at ambient pressure at different temperatures (indicated) is shown in Fig.2.

\begin{figure}
\includegraphics[angle=0,width=0.45\textwidth]{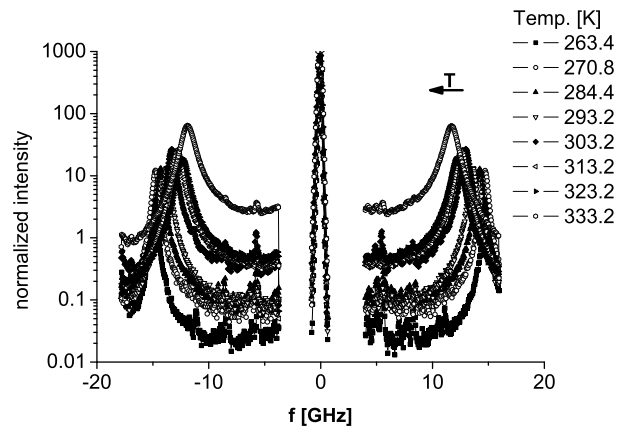}
\caption{\label{Fig.2}Temperature dependence of the polarized (VV) Brillouin spectra of BMMPC measured in the back scattering geometry.}
\end{figure}

The central (attenuated) part of the spectra represents the instrumental (resolution) function of the Tandem FPI. It can be clearly seen that the frequency of the maximum is decreasing and the width of the Brillouin line is increasing with increasing temperature. 

Similar measurements were performed for BMPC and OTP. Also the pressure dependence of the Brillouin spectra was studied for selected temperatures. The dependence of the half width at half height $\Gamma_B$ of the longitudinal Brillouin lines at selected pressures is plotted versus the distance from $T_g: T-T_g$, in Fig.3.  
As it is evident from Fig.3, the width of the Brillouin peak for BMPC is always larger than that for BMMPC, even after correction for the difference between $T_g$'s. Additionally, at the glass transition temperatures, $T_g$, the value of $\Gamma_B$ is still substantial and amounts to about 0.2 GHz for BMPC and 0.09 GHz for BMMPC, respectively.

\begin{figure}
\includegraphics[angle=0,width=0.45\textwidth]{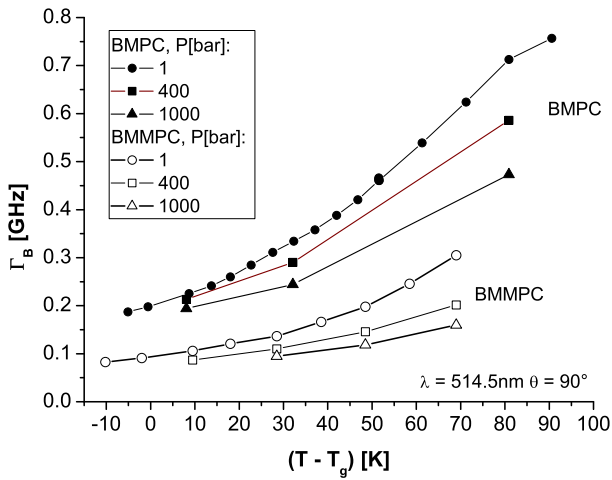}
\caption{\label{Fig.3}Temperature dependence of the width of longitudinal Brillouin lines at selected pressures (indicated) for BMPC and BMMPC.}
\end{figure}

It has been shown earlier \cite{Gerd, Kahle}, that in BMPC two internal relaxations are present: the phenyl flip and rotation of the O-CH$_3$ group, while in the chemically very similar BMMPC the phenyl flip is no longer possible and is replaced by low amplitude librational motions of the rings. Thus, this behavior of $\Gamma_B$ for BMPC can be attributed to the presence of a strong internal relaxation - the phenyl flip in this molecule. In the case of BMMPC a weaker librational dynamics leads to a much smaller broadening of the Brillouin line. Obviously, the structural relaxation (density fluctuations) in these supercooled liquids can also contribute to the damping of the longitudinal phonons and broadening of the Brillouin peaks \cite{CD}. In order to separate the contributions of the structural relaxation and internal dynamics to the $\Gamma_B$, the pressure dependence of the Brillouin spectra was also measured.

As we can see in Fig.3, close to the glass transition $T_g$ the increasing pressure is not changing the $\Gamma_B$ value very much. Additionally, the pressure dependence of $\Gamma_B$ is the strongest at the highest temperature and very weak at the lowest temperature close to $T_g$. This behavior can be explained in the following way: The high pressure value of the $\Gamma_B(P)$ corresponds to the contribution of the internal relaxations, since these processes should not depend on pressure (in this pressure range) but only on temperature. The contribution of the structural relaxation to the $\Gamma_B(P)$ can be estimated from the difference between the value of $\Gamma_B(P)$ at a given pressure and the high pressure value. This contribution is the highest at the highest temperature and the lowest pressure studied where the structural relaxation time is the shortest and thus the corresponding phonon damping is most efficient.

Similar behavior can be observed for BMPC. The main difference is that the value of $\Gamma_B$ at high pressure and at the lowest temperature close to $T_g$ is approximately a factor of two higher than that for BMMPC. This difference may be attributed to the presence of an internal relaxation - the phenyl flip which in BMMPC is replaced by a weaker librational process.

A combined lin-log plot of the width of Brillouin lines corresponding to the longitudinal and transverse phonons versus $T-T_g$ is shown in Fig.4, where also the corresponding values for OTP are show.

\begin{figure}
\includegraphics[angle=0,width=0.45\textwidth]{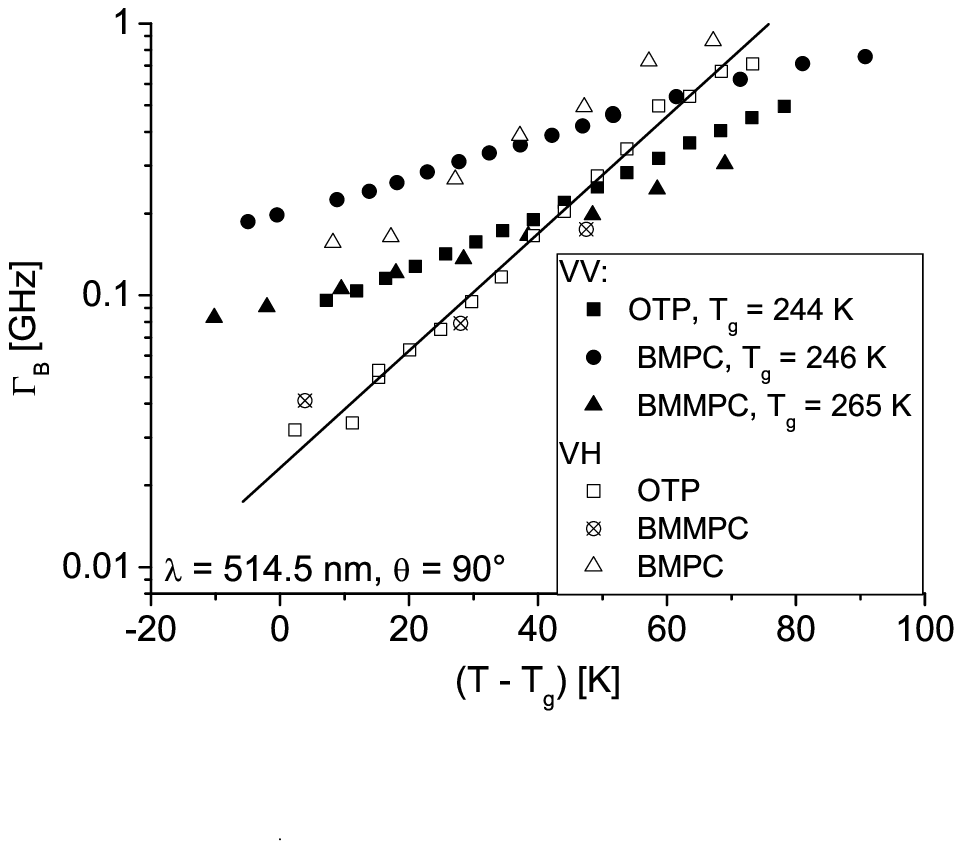}
\caption{\label{Fig.4}The half width at half height $\Gamma_B$ of longitudinal (VV) and transverse (VH) Brillouin lines measured for OTP, BMPC and BMMPC at ambient pressure.}
\end{figure}

The OTP values were re-measured by us and are in a very good agreement with the literature data \cite{Steffen1, AP, Monaco2, Wang, Steffen2}. As one can see in this figure, the width of the longitudinal and transverse Brillouin lines for BMMPC and OTP behave according to the predictions of the Zwanzig model: The width of the longitudinal Brillouin lines at $T_g$ is substantially different from zero while the width of transverse Brillouin lines approaches zero at $T_g$. This can be explained by the presence of weak internal relaxation which, in the limit of weak coupling of internal dynamics to translational motion, results in an additional damping (broadening) of longitudinal phonons (Brillouin lines), leaving the transverse phonons (and the frequency dependent shear viscosity) unaffected.

The situation is completely different in the case of BMPC. There, the width of both longitudinal and transverse Brillouin lines at $T_g$ is substantially different from zero, so obviously, the weak coupling limit and the Zwanzig model do not hold. Additionally, the excess width at $T_g$ is not related to the structural ($\alpha$-) relaxation because, as it will be shown in detail elsewhere \cite{full paper} it does not depend on pressure. Thus, the only reason for the unusual behaviour of BMPC, not observed so far in any liquid, is the strong internal relaxation, i.e. the phenyl flip which is present in this molecule, while in BMMPC and OTP only much weaker internal relaxations, of the form of librational and stretching motions of the rings, are present. Thus, we can see, that in the case of molecules having strong internal relaxations, the width of both longitudinal and transverse Brillouin lines is additionally broadened, beyond the usual width due to the structural relaxation, in contrast to the predictions of the Zwanzig model.

We conclude, that in liquids composed of molecules with internal degrees of freedom, the frequency dependent shear viscosity and the width (damping) of transverse Brillouin lines (phonons) can be affected by internal modes. This is a theoretically not predicted new effect which occurs together with the usual effect of internal modes on the frequency dependent bulk viscosity and damping of longitudinal phonons, explained theoretically by Zwanzig.

Pressure dependent measurements of the polarized and depolarized Brillouin spectra allow to separate the contributions of structural relaxation and intramolecular processes to the width (damping) of both longitudinal and transverse Brillouin lines (phonons). 

Theoretical model developed by Zwanzig for the limiting case of weak coupling of internal relaxations to the translational degrees of freedom does not describe this new effect, as already shown by the author \cite{Zwanzig}. A more general theory is needed, which would cover also the case of strong coupling and other possible mechanisms in order to explain the effect of internal dynamics on the frequency dependent shear viscosity and broadening (damping) of transverse Brillouin lines (phonons).

\begin{acknowledgments}
Partial financial support from the Polish Ministry of Scientific Research and Information Technology (grant No. 1 PO3B 083 26) is greatfully acknowledged. A.P. and C.D. acknowledge the support of Polish-French cooperation grant "Polonium" (No. 075546UE).
\end{acknowledgments}

% Create the reference section using BibTeX:
%\bibliography{basename of .bib file}

\begin{thebibliography}{99}
\bibitem{Zwanzig} R. Zwanzig, {J. Chem. Phys.} {\bf 43}, 714 (1965).
\bibitem{Monaco2} G. Monaco, S. Caponi, R. DiLeonardo, D. Fioretto and G. Ruocco, {Phys. Rev. E} {\bf 62}, R7595 (2000).
\bibitem{Mountain} R.D. Mountain, {J. Res. NBS (Phys. and Chem.)} {\bf 72A}, 95 (1968).
\bibitem{Steffen1} W. Steffen, A. Patkowski, G. Meier, E.W. Fischer, {J. Chem. Phys.} {\bf 96}, 4171, (1992).
\bibitem{AP} A. Patkowski, W. Steffen, G. Meier, E.W. Fischer, {J. Non-Crystalline Solids}, {\bf 172-174}, 52 (1994).
\bibitem{Monaco1} G. Monaco, D. Fioretto, C. Masciovecchio, G. Ruocco and F. Sette, {Phys. Rev. Lett.} {\bf 82}, 1776 (1999).
\bibitem{Mossa} S. Mossa, G. Monaco and G. Ruocco, {Europhys. Let.} {\bf 60}, 92 (2002).
\bibitem{Gerd} G. Meier, B. Gerharz, D. Boese, and E.W. Fischer, {J. Chem. Phys.} {\bf 94}, 3050 (1991).
\bibitem{Gerharz} B. Gerharz, G. Meier, and E.W. Fischer, {J. Chem. Phys.} {\bf 92}, 7110 (1990).
\bibitem{Kahle} S. Kahle, J. Gapinski, G. Hinze, A. Patkowski, and G. Meier, {J. Chem. Phys.} {\bf 122}, 074506 (2005).
\bibitem{HPcell} G. Fytas, Th. Dorfmueller, and C. H. Wang, {J. Phys. Chem.} {\bf 87}, 5045 (1983).
\bibitem{CD} C. Dreyfus, A. Le Grand, J. Gapinski, W. Steffen and A. Patkowski, {Eur. Phys. J. B} {\bf 42}, 309 (2004).
\bibitem{Wang} Y. Higashigaki and C.H. Wang, {J. Chem. Phys.} {\bf 74}, 3175 (1981); C.H. Wang, X.R. Zhu, J.C. Shen, {Mol. Phys.} {\bf 62}, 749 (1987).
\bibitem{Steffen2} W. Steffen, B. Zimmer, A. Patkowski, G. Meier, E.W. Fischer, {J. Non-Crystalline Solids}, {\bf 172-174}, 37 (1994).
\bibitem{full paper}  unpublished
\end{thebibliography}

\end{document}